\documentclass[aps,prb,superscriptaddress,twocolumn,showpacs]{revtex4}
\usepackage{graphicx}
\usepackage{amsmath}
\usepackage{amssymb}
\usepackage{amsfonts}

\begin{document}

\title{Anomalous Josephson current through a ferromagnetic trilayer junction}
\author{Jun-Feng Liu}
\author{K. S. Chan}
\email{apkschan@cityu.edu.hk}
\affiliation{Department of Physics and Materials Science, City University of Hong Kong,
Tat Chee Avenue, Kowloon, Hong Kong, People's Republic of China}

\begin{abstract}
We studied the anomalous Josephson current appearing at zero phase
difference in junctions coupled with a ferromagnetic trilayer which has
noncoplanar magnetizations. A $\pi/2$ junction with an equilibrium phase
difference $\pi/2$ is obtained under suitable conditions. The equilibrium
phase difference and the amplitude of the supercurrent are all tunable by
the structure parameters. In addition to calculating the anomalous current
using the Bogoliubov-de Gennes equation, we also developed a clear physical
picture explaining the anomalous Josephson effect in the structure. We show
that the triplet proximity correlation and the phase shift in the anomalous
current-phase relation all stem from the spin precession in the first and
third ferromagnet layers.
\end{abstract}

\maketitle


\section{Introduction}

Usually the supercurrent in a Josephson junction vanishes, when the phase
difference between the two superconductors is zero, and in the tunneling
limit the current-phase relation (CPR) is sinusoidal $I(\varphi )=I_{c}\sin
(\varphi )$. \cite{golubov04} Recently some studies \cite%
{krive1,feinberg,buzdin08,buzdin09,martin09,braude07,grein09,tanaka09} found
an anomalous Josephson current flow $I_{a}$ exists even at zero phase
difference ($\varphi =0$). The anomalous supercurrent is equivalent to the
presence of an additional phase shift $\varphi _{0}$ in the conventional
CPR, i.e., $I(\varphi )=I_{c}\sin (\varphi +\varphi _{0})$. In fact, such
CPRs have been predicted for Josephson junctions of unconventional
superconductors, \cite{Geshkenbein,Yip,Sigrist,Tanaka,brydon08} but the
experimental verification is still lacking. Recent studies have shown that
the anomalous supercurrent can also exist in junctions with conventional
s-wave BCS superconductors if both spin-orbit interaction (SOI) and a
suitably oriented Zeeman field are present in the coupling layer. \cite%
{krive1,feinberg,buzdin08,buzdin09,martin09} These studies revealed that the
anomalous effect in conventional junctions has some intricate physics. More
interesting, an anomalous Josephson current can also appear in
superconductor (S)-ferromagnet(F) hybrid structure without SOI. \cite%
{braude07,grein09} In Grein's study, \cite{grein09} a SFS hybrid structure
with two spin-active interfaces was considered. The two spin-active
interfaces are critical to the triplet proximity effect and the anomalous
supercurrent in the structure, but the physics is still unclear.

In this study, we generalize the two spin-active interfaces to two
ferromagnetic layers with finite thicknesses and clarify the physical
mechanisms responsible for the anomalous supercurrent. In such SFFFS
structures, we find that the triplet proximity correlation and the phase
shift in the anomalous CPR all stem from the spin precession in the first
and third F layers. According to the symmetry analysis, \cite{liu10} an
anomalous supercurrent is possible when the symmetries of the time-reversal
operator $T$ and its combination with a spin rotation operator with respect
to an arbitrary spin quantum axis $\mathbf{n}$ $\sigma _{\mathbf{n}}T$ are
broken at the same time. As a result, the simplest superconductor
(S)-ferromagnet (F)-superconductor (S) junction for achieving an anomalous
Josephson current requires the F layer to be a ferromagnetic trilayer with
noncoplanar magnetizations for breaking the symmetry of the operator $\sigma
_{\mathbf{n}}T$. SFFFS junctions where the magnetizations of the three
ferromagnetic layers need not be noncoplanar \cite{braude07,houzet07} and
SFS junctions with inhomogeneous magnetization \cite%
{bergeret01,bergeret012,jonson01,linder10}, have been studied in
order to understand the effects of triplet correlation induced in
the F layers. Controllable $0$-$\pi$ transition and spin-triplet
supercurrents have been realized experimentally recently
\cite{robinson101,robinson102}. In our study, we found that
triplet correlation is also an important condition for the
anomalous supercurrent.
\cite{eschrig07,eschrig08,tanaka07,volkov10,luka10}

\begin{figure}[btp]
\begin{center}
\includegraphics[bb=16 16 307 235, width=3.105in]{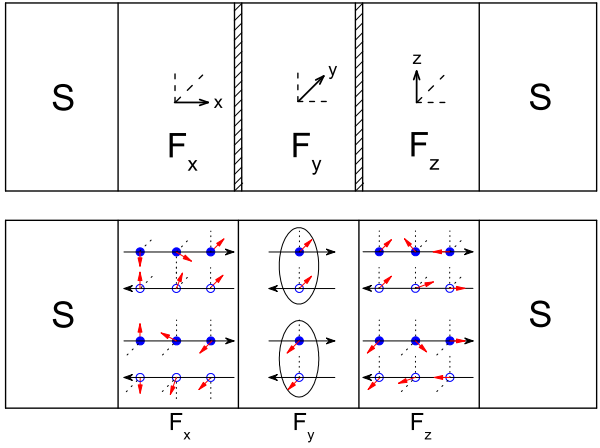}
\end{center}
\caption{(Color online) Upper panel: Schematic diagram
of the $SF_xF_yF_zS$ junction where two barriers are present
between the F layers. Lower panel: Schematic illustration of the
formation of Andreev bound states with triplet correlation in the
$F_y$ layer due to spin precession of electrons and holes in the
$F_x$ and $F_z$ layers.} \label{model}
\end{figure}

We consider a junction consisting of two conventional s-wave superconductors
coupled by a ferromagnetic trilayer with noncoplanar magnetizations. For
convenience, hereafter we denote the three F layers sequentially by $F_1,
F_2, F_3$. We start with the typical situation where the magnetizations are
along the $x$, $y$, $z$ axes respectively (i.e. an $SF_xF_yF_zS$ junction),
as shown in the upper panel of Fig. \ref{model}. This junction is a $\pi/2$
junction with an equilibrium phase difference $\pi/2$ under suitable
conditions. The equilibrium phase difference can be tuned by the lengths,
the exchange energies, and the magnetization orientations of the $F_1$ and $%
F_3$ layers. And the amplitude of the supercurrent can be tuned by the
barriers between the F layers or by the length and the exchange energy of
the middle $F_2$ layer. In this regime the Josephson junction can also act
as a supercurrent rectifier. \cite{zapata96,carapella01}

The paper is organized as follows. In Sec. II we present the model and solve
the scattering problem for quasi-particles based on the Bogoliubov-de Gennes
equation. The Josephson current and Andreev bound states can be obtained
from the scattering matrices. In Sec. III we show the numerical results for
the anomalous supercurrent and corresponding Andreev bound states and reveal
the physics. A conclusion and remarks will be given in Sec. IV.

\section{Model and formalism}

In the numerical calculation, we consider $SF_{1}F_{2}F_{3}S$ junctions with
various lengths and exchange energies for each F layer and various barrier
strengths for the two barriers between the F layers. The transport direction
is along the $x$ axis. The three F layers have the thicknesses, $%
L_{1},L_{2},L_{3}$, the exchange energies, $h_{1},h_{2},h_{3}$, and the
magnetization orientations, $(\theta _{1},\phi _{1}),(\theta _{2},\phi
_{2}),(\theta _{3},\phi _{3})$ in spherical coordinates. The effective
Hamiltonian of the system is given by \cite{yokoyama06,tinkham82}
\begin{equation}
H=\left(
\begin{array}{cccc}
\epsilon _{k}+h_{z} & h_{xy}^{\ast } & 0 & \Delta (x) \\
h_{xy} & \epsilon _{k}-h_{z} & -\Delta (x) & 0 \\
0 & -\Delta ^{\ast }(x) & -\epsilon _{k}-h_{z} & -h_{xy} \\
\Delta ^{\ast }(x) & 0 & -h_{xy}^{\ast } & -\epsilon _{k}+h_{z}%
\end{array}%
\right)
\end{equation}%
where $\epsilon _{k}=\frac{\hbar ^{2}}{2m}(k_{x}^{2}+k_{y}^{2}-k_{F}^{2})+U$
with $k_{F}$ the Fermi wave number, $U=U_{0}[\delta (x-L_{1})+\delta
(x-L_{1}-L_{2})]$ represents the two barriers between the F layers, and $%
h_{z}=h\cos \theta $, $h_{xy}=h\sin \theta e^{i\phi }$, with $h$ the
strength and $(\theta ,\phi )$ the orientation of the exchange field; $%
\Delta (x)=\Delta \lbrack \Theta (-x)e^{i\varphi /2}+\Theta
(x-L)e^{-i\varphi /2}]$ describes the pair potential with $%
L=L_{1}+L_{2}+L_{3} $ and $\Delta $ the bulk superconducting gap and $%
\varphi =\varphi _{L}-\varphi _{R}$ the macroscopic phase difference of the
two superconductor leads. The temperature dependence of the magnitude of $%
\Delta $ is given by $\Delta (T)=\Delta (0)\tanh (1.74\sqrt{T_{c}/T-1})$.
\cite{muh59} Since the transversal momentum components are conserved and not
important to the total Josephson current, we consider the question in the
one-dimensional regime for simplicity. The Bogoliubov-de Gennes equation can
be easily solved for each superconductor lead and each F layer respectively.
The scattering problem can be solved by considering the boundary conditions
at the interfaces. Each interface gives a scattering matrix. The total
scattering matrix of the system can be obtained by the combination of all
these scattering matrices of interfaces. From the total scattering matrix,
we can obtain the Andreev reflection amplitudes $a_{1\sigma }$ and $%
a_{2\sigma }$ of the junction where $a_{1\sigma }$ is for the reflection
from an electron-like to a hole-like quasiparticle and $a_{2\sigma }$ is for
the reverse process with $\sigma $ representing the spin. The stationary
Josephson current can be expressed in terms of the Andreev reflection
amplitudes by using the temperature Green function formalism \cite{ft91}
\begin{equation}
I_{e}(\varphi )=\frac{e\Delta }{4\hbar }\sum_{\omega _{n},\sigma }\frac{%
k_{B}T}{\Omega _{n}}(k_{n}^{+}+k_{n}^{-})\left( \frac{a_{1\sigma n}}{%
k_{n}^{+}}-\frac{a_{2\sigma n}}{k_{n}^{-}}\right) ,  \label{current}
\end{equation}%
where $k_{n}^{+}$, $k_{n}^{-}$, $a_{1\sigma n}$, and $a_{2\sigma n}$ are
obtained from $k_{s}^{+}$, $k_{s}^{-}$, $a_{1\sigma }$, and $a_{2\sigma }$
by analytic continuation $E\rightarrow i\omega _{n}$. $k_{s}^{\pm }$ is the
wave vector for electron or hole in the superconductors and the Matsubara
frequencies are $\omega _{n}=\pi k_{B}T(2n+1)$, $n=0,\pm 1,\pm 2,\cdots $,
and $\Omega _{n}=\sqrt{\omega _{n}^{2}+\Delta ^{2}}$.

The discrete spectrum of the Andreev bound states can be determined by using
the condition \cite{beenakker91}
\begin{equation}
\det [1-R_{2}PR_{1}P]=0
\end{equation}%
where $R_{1},R_{2},P$ are $4\times 4$ matrices, $P$ is the propagation
matrix of modes in the $F_{2}$ layer, and $R_{1}$ ($R_{2}$) is the
reflection matrix of the right-going (left-going) incident waves.

In order to study the spin properties of the Andreev bound states formed at $%
F_{2}$ layer, we can also work out the Green's function
$G(x,x^{\prime },E)$ in $F_{2}$ layer which is a $4\times 4$
matrix. \cite{zaikin09} Now it is convenient to take the eigen
spinors of $F_{2}$ layer, i.e., spin-parallel and
spin-antiparallel with respect to the exchange field
$\mathbf{h}_{2}$ as the unit vectors of the
spin space. Then the spin current in $F_{2}$ layer can be evaluated by \cite%
{linder10}
\begin{eqnarray}
I_{s}(\varphi ) &=&\frac{\hbar ^{2}k_{B}T}{4mi}\lim_{x^{\prime }\rightarrow
x}\left( \frac{\partial }{\partial x^{\prime }}-\frac{\partial }{\partial x}%
\right) \sum_{\omega _{n}}  \notag \\
&&Tr\left\{ \left(
\begin{array}{cc}
\sigma _{z} & 0 \\
0 & \sigma _{z}%
\end{array}%
\right) G_{\omega _{n}}(x,x^{\prime })\right\}  \notag \\
&=&\frac{\hbar }{2e}(I_{+}-I_{-}),
\end{eqnarray}%
where $I_{+}$ ($I_{-}$) is the charge currents of electrons with parallel
spin (antiparallel spin) and obviously satisfies $I_{e}=I_{+}+I_{-}$.

\section{Results and discussion}

We start with the typical noncoplanar magnetization configuration, i.e., the
$SF_xF_yF_zS$ junction. For simplicity, we introduce the dimensionless
units: the energy $E\rightarrow EE_{F}$, the wave vector $\mathbf{k}%
\rightarrow\mathbf{k}k_{F}$, the coordinate $\mathbf{x}\rightarrow\mathbf{x}%
/k_{F}$, and the strength of exchange field $h\rightarrow hE_{F}$. All
physical quantities are expressed in the dimensionless units in the rest of
the paper. The superconductors considered are characterized with $%
\Delta=10^{-3}$ which corresponds to the BCS coherence length at zero
temperature $\xi_{0}=2/\pi\Delta\approx636.6$.

\begin{figure}[btp]
\begin{center}
\includegraphics[bb=18 16 233 232, width=3.205in]{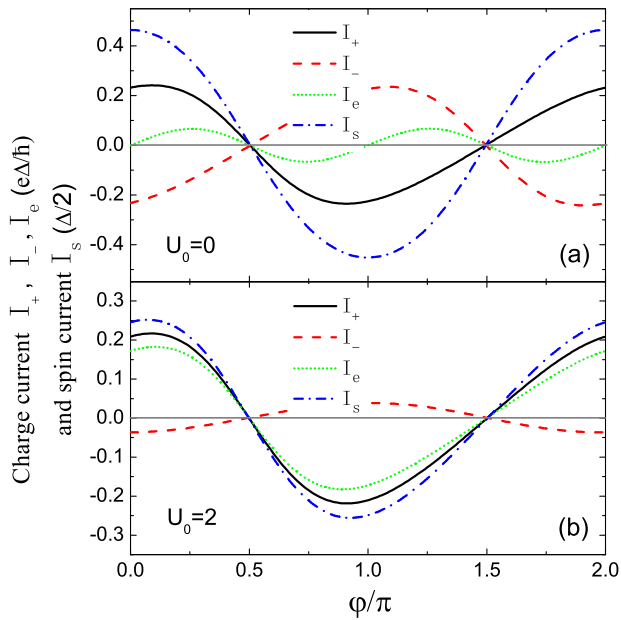}
\end{center}
\caption{(Color online) Charge and spin currents
$I_+$, $I_-$, $I_e$, $I_s$ versus $\protect\varphi$ for the
$SF_xF_yF_zS$ junction. The barrier strength $U_0=0$ for (a) and
$U_0=2$ for (b). The strength of exchange
fields $h_1=h_2=h_3=0.05$, and the lengths of F layers $L_1=L_2=L_3=10%
\protect\pi$. The temperature $T/T_{c}=0.5$ with $T_{c} $ being the critical
temperature.}
\label{is}
\end{figure}

\begin{figure}[btp]
\centering
\includegraphics[bb=16 16 290 230, width=3.205in]{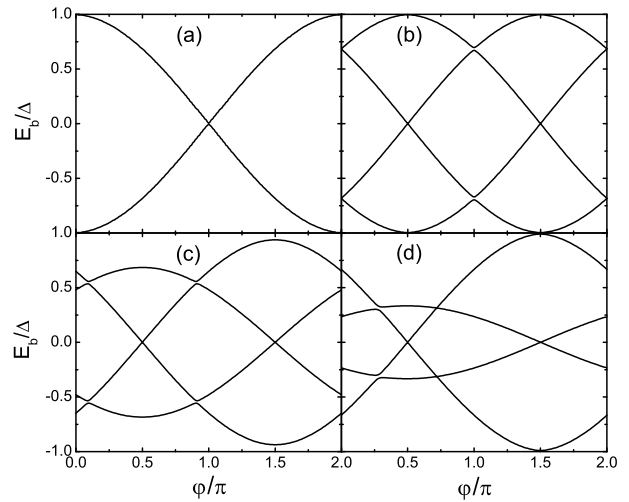}
\caption{The energy levels of the Andreev bound states $E_b$. The strength
of exchange fields $h_1=h_2=h_3=0$ for (a) and $h_1=h_2=h_3=0.05$ for (b),
(c), (d). The barrier strength $U_0=0$ for (b), $U_0=1$ for (c), and $U_0=2$
for (d). The other parameters are the same as those in Fig. \protect\ref{is}%
. }
\label{abs}
\end{figure}

Fig.~\ref{is} shows the charge and spin currents $I_{+}$, $I_{-}$, $I_{e}$, $%
I_{s}$ as functions of the phase difference $\varphi $ for the $%
SF_{x}F_{y}F_{z}S$ junction. The corresponding Andreev bound states are
shown in Fig. \ref{abs}. It is interesting to note that when there is a
barrier between the F layers, ($U_0=2$), there is a significant anomalous
Josephson current. When there is no barrier $U_0=0$, the anomalous Josephson
current is nearly zero. This interesting dependence on the barrier strength $%
U_0$ will be explained below in terms of the spin characteristics of the
Andreev bound states in the $F_{y}$ layer.

Firstly, it is useful to point out a large spin current exists in the $F_{y}$
layer, implying that the superconductivity correlation is mainly triplet in
the $F_{y}$ layer. This is easily understood by considering the formation of
an Andreev bound state in the $F_{y}$ layer. A right-going electron with
spin parallel to the $y$ axis $(1,i)^{T}$ from the $F_{y}$ layer will have
its spin precessing about the $z$ axis in the $F_{z}$ layer before it
reaches the right superconductor. After the Andreev reflection from the
right superconductor, a hole with reverse spin goes left and its spin
continues to precess. The one-way angle of precession is approximately $%
(k_{+}-k_{-})L_{3}\approx h_{3}L_{3}$ where $k_{+}$ ($k_{-}$) is the
wave-vector of up-spin (down-spin ) quasi-particle. Thus if the condition $%
h_{3}L_{3}=n\pi +\pi /2$ ($n$ is an integer) is satisfied, the reflected
hole from the right superconductor will have its spin parallel to the
incident electron's spin in the $F_{y}$ layer. An Andreev bound state is
formed, after this reflected hole travels through the $F_y$ and $F_x$ layers
and Andreev reflected from the left superconductor and changes to an
electron to move right to finish a cycle. If the spin rotation angle in the $%
F_{x}$ layer satisfies the same condition $h_{1}L_{1}=n\pi +\pi /2$. The
electrons and holes have identical spins (parallel to the $y$ axis) in the $%
F_y$ layer and the Andreev bound state formed has complete triplet
correlation in the $F_{y}$ layer, as schematically shown in the
lower panel of Fig. \ref{model}. Triplet correlation can exist in
other different type of magnetic inhomogeneity too. Bergeret
\textit{et al.} \cite{bergeret01} have studied S/F/S junctions
with spiral magnetization in the F layer and found spin triplet
correlation there. In the present model, we found two Andreev
bound states below the Fermi level with complete triplet
correlation; one is "spin-up" (with respect to the $y$ axis),
which carries the current $I_{+}$, and one
is "spin-down", which carries the current $I_{-}$, as shown in Fig. \ref{abs}%
. In the short junction limit, the Josephson current is totally carried by
the Andreev bound states. \cite{zhu96}

Besides complete triplet correlation in the $F_{y}$ layer, another
interesting feature noted in Fig.~\ref{is} is that $I_{+}$ has a phase shift
of $\pi /2$ while $I_{-}$ has a phase shift of $-\pi /2$ compared with the
conventional CPR. So, these two currents move in opposite directions. Now we
follow the Andreev reflection processes occurring in the formation of the
bound states to find out the phase shift. For simplicity, we assume $%
h_{1}=h_{2}=h_{3}=h\,$; thus, the wave-vectors of "spin-up" ($+$) and
"spin-down" ($-$) electrons (holes) with energy $E$ at each F layer are $%
k_{\pm }^{e(h)}=\sqrt{k_{F}^{2}+\rho _{e(h)}E\mp h}$ with $\rho
_{e(h)}=+(-)1 $. In the short junction limit and the limit $E\ll h\ll E_{F}$%
, we have $k_{\pm }^{e}\approx k_{\pm }^{h}\approx k_{\pm }=k_{F}\mp \frac{h%
}{2}$. We start with a right-going "spin-up" electron at the position $%
x=L_{1}+0$, the wave function can be written as $(1,i,0,0)^{T}$. The
electron moves right and acquires a phase $e^{ik_{+}L_{2}}$ when it arrives
at the interface $x=L_{1}+L_{2}$. To simplify the discussion we focus on the
Andreev reflections at the F/S interfaces and ignore the normal reflections
at the barriers which affect only the amplitude of the supercurrent but not
the phase shift. When the electron travels through the $F_{z}$ layer, its
spin precesses. The state becomes $%
(e^{ik_{+}L_{3}},ie^{ik_{-}L_{3}},0,0)^{T}e^{ik_{+}L_{2}}$ when the electron
arrives at the interface $x=L_{1}+L_{2}+L_{3}$. Then, the electron is
reflected as a hole with reverse spin and the hole wave function is $%
(0,0,-ie^{ik_{-}L_{3}},e^{ik_{+}L_{3}})^{T}e^{ik_{+}L_{2}}\frac{v}{u}%
e^{i\varphi /2}$ where $u=\sqrt{(1+\Omega /E)/2}$, $v=\sqrt{(1-\Omega /E)/2}$
with $\Omega =\sqrt{E^{2}-\Delta ^{2}}$. The algebraic derivation is not
shown here for space limitation and the approximation $k_{\pm }^{e}\approx
k_{\pm }^{h}\approx k_{\pm }^{s}\approx k_{F}$ has been used in the
derivation where $k_{+}^{s}$ ($k_{-}^{s}$) is the wave-vector of
electronlike (holelike) quasiparticle in the superconductors. The
Andreev-reflected hole moves left and has its spin rotated in the $F_{z}$
layer again and then goes back to the $F_{y}$ layer $x=L_{1}+L_{2}-0$. Now
the wave function becomes $%
(0,0,-ie^{ihL_{3}},e^{-ihL_{3}})^{T}e^{ik_{+}L_{2}}\frac{v}{u}e^{i\varphi
/2}=(0,0,1,-i)^{T}e^{ik_{+}L_{2}}\frac{v}{u}e^{i\varphi /2}$ where the
condition $hL_{3}=\pi /2$ has been used. The wave function describes a
"spin-up" hole with respect to the $y$ direction. Then the hole goes left
through the $F_{y}$ layer and acquires a phase $e^{-ik_{+}L_{2}}$. So the
wave function becomes $(0,0,1,-i)^{T}\frac{v}{u}e^{i\varphi /2}$ when the
hole arrives at the interface $x=L_{1}$. Consequently, the hole has its spin
precessed in the $F_{x}$ layer and moves left to the interface $x=0$ with
the wave function $\frac{1}{2}\left[ (1-i)(0,0,1,1)^{T}e^{-ik_{+}L_{1}}%
\right. +\left. (1+i)(0,0,1,-1)^{T}e^{-ik_{-}L_{1}}\right] \frac{v}{u}%
e^{i\varphi /2}$. The hole is Andreev-reflected as an electron with reverse
spin described by $\frac{1}{2}\left[ (1-i)(1,-1,0,0)^{T}e^{-ik_{+}L_{1}}%
\right. -\left. (1+i)(1,1,0,0)^{T}e^{-ik_{-}L_{1}}\right] \left( \frac{v}{u}%
\right) ^{2}e^{i\varphi }$. Then the electron goes through the $F_{x}$ layer
again and back to the starting position $x=L_{1}+0$ to finish a cycle. The
final wave function is $(1,i,0,0)^{T}e^{i\pi /2}\left( \frac{v}{u}\right)
^{2}e^{i\varphi }$ where $hL_{1}=\pi /2$ is used. Comparing with the initial
wave function $(1,i,0,0)^{T}$, we can see the phase shift of the "spin-up"
Andreev bound state is indeed $\pi /2$ when considering a conventional CPR.
In the same way, we can find out the phase shift of the "spin-down" Andreev
bound state is $-\pi /2$. In this round-trip cycle of the quasi-particle, we
can clearly see that the phase shifts in Andreev bound states come from the
spin precession of electron and hole in the $F_{x}$ and $F_{z}$ layers.

\begin{figure}[btp]
\centering
\includegraphics[bb=18 14 305 219, width=3.405in]{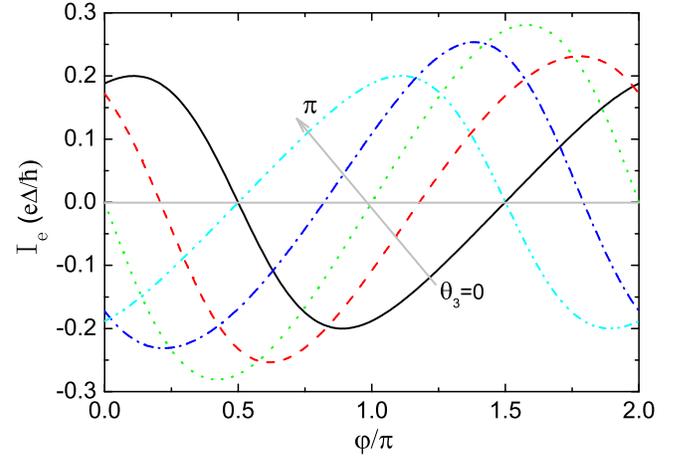}
\caption{(Color online) $I_e$ versus $\protect\varphi$ for the $SF_xF_yF_zS$
junction with different $\protect\theta_3$: varying $\protect\theta_3$ from $%
0$ to $\protect\pi$ with a step $\protect\pi/4$, $\protect\phi_3=0$. The
other parameters are chosen as: $U_0=2$, $h_1=h_2=h_3=0.1$, $L_1=L_2=L_3=5%
\protect\pi$, $T/T_{c}=0.5$.}
\label{t3}
\end{figure}

If we neglect the second-harmonic term in the CPR, the charge current
carried by the two Andreev bound states can be written as \cite{buzdin05}%
\begin{equation}
I_{+}\approx I_{+}^{0}\sin (\varphi +\frac{\pi }{2}),\text{ \ }I_{-}\approx
I_{-}^{0}\sin (\varphi -\frac{\pi }{2}),
\end{equation}%
where $I_{+}^{0}$ ($I_{-}^{0}$) is the amplitude of the "spin-up"
("spin-down") charge current. When the barriers are absent, the normal
scattering at the two F/F interfaces can be ignored and we can have $%
I_{+}^{0}\approx I_{-}^{0}$. As a result, the total charge current $%
I_{e}=I_{+}+I_{-}$ is very small and only the second-harmonic term remains,
as shown in Fig.~\ref{is} (a). At zero phase difference, the charge current
is very small and the spin current in the $F_{y}$ layer is almost a pure
spin current.

When the barriers are present, the normal scattering at the barriers reduces
the amplitudes of the two charge currents $I_{+}^{0}$ and $I_{-}^{0}$. \ The
transmission probability through the double delta function barriers of
electrons or holes depends on the wave-vector of the particle in the $F_y$
layer and reaches the maximum when resonance transmission occurs. Here in
the $F_{y}$ layer, the "spin-up" Andreev bound state couples a "spin-up"
electron with a "spin-up" hole which have the same wave-vector $k_{+}\approx
k_{F}-\frac{h}{2}$ while the "spin-down" Andreev bound state couples a
"spin-down" electron with a "spin-down" hole which have the same wave-vector
$k_{-}\approx k_{F}+\frac{h}{2}$. The difference in the wave vector between
the two Andreev bound states leads to the difference in the transmissions
through the $F_y$ layer. Consequently, we can make a large difference
between $I_{+}^{0}$ and $I_{-}^{0}$ as shown in Fig.~\ref{is} (b) by using
two barriers as well as suitable exchange field strength and length of the $%
F_{y}$ layer. In this way, an anomalous Josephson current appears at zero
phase difference. The CPR of the junction has a phase shift of $\pm \pi /2$
in comparison with the conventional CPR where the sign of the phase shift
depends on the relative magnitude of $I_{+}^{0}$ and $I_{-}^{0}$.

\begin{figure}[btp]
\centering
\includegraphics[bb=18 16 313 150, width=3.405in]{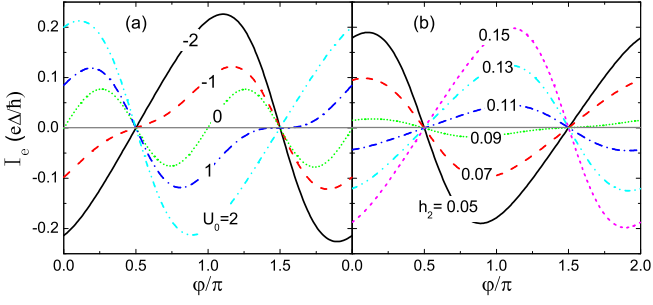}
\caption{(Color online) $I_e$ versus $\protect\varphi$ for the $SF_xF_yF_zS$
junction with different $U_0$ and $h_2$: (a) varying $U_0$ from $-2$ to $2$
with a step $1$, $h_2=0.1$, $L_2=5\protect\pi$; (b) varying $h_2$ from $0.05$
to $0.15$ with a step $0.02$, $L_2=10\protect\pi$, $U_0=2$. The other
parameters are chosen as: $h_1=h_3=0.1$, $L_1=L_3=5\protect\pi$, $%
T/T_{c}=0.5 $.}
\label{f2}
\end{figure}

\begin{figure}[tbp]
\centering
\includegraphics[bb=18 37 327 177, width=3.405in]{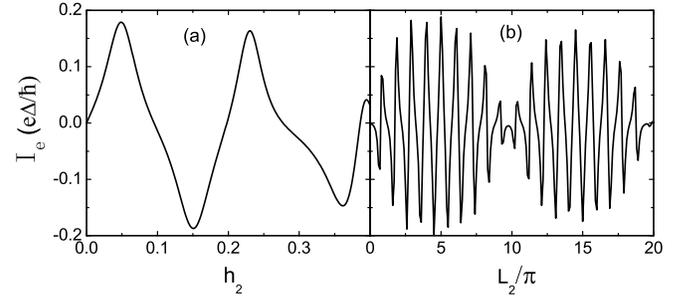}
\caption{The anomalous supercurrent at zero phase difference $I_{e}(\protect%
\varphi =0)$ for the $SF_{x}F_{y}F_{z}S$ junction as functions of: (a) $%
h_{2} $ with $L_{2}=10\protect\pi $, and (b) $L_{2}$ with $h_{2}=0.1$. The
other parameters are chosen as: $h_{1}=h_{3}=0.1$, $L_{1}=L_{3}=5\protect\pi
$, $U_{0}=2$, $T/T_{c}=0.5$.}
\label{h2l2}
\end{figure}

Since the phase shift of the anomalous CPR stems from the spin precession of
electrons and holes in the $F_{x}$ and $F_{z}$ layers, we can modulate the
phase shift by tuning the parameters of these two layers. If the conditions $%
\phi _{1}=\phi _{3}=0$ and $h_{1}L_{1}=h_{3}L_{3}=(n+1/2)\pi $ are
satisfied, the complete equal-spin triplet correlation in the
$F_{y}$ layer is maintained. Now the phase shifts of the "spin-up"
and "spin-down" Andreev bound states are $\pm (\pi +\theta
_{3}-\theta _{1})$ according to the above discussion. Fig.
\ref{t3} shows the tuning of the equilibrium phase difference by
varying $\theta _{3}$. Bergeret \textit{et al.} \cite{bergeret01}
have found that the relative orientation of the two magnetizations
in S/F/I/F/S junctions can change the critical current.  For the
particular structure considered in Fig. \ref{t3}, which is
different from theirs, the orientation of the magnetization in the
third layer has no strong effect on the critical current. However,
for other values of $h_2$, the orientation can also modify the
critical current. On the other hand, the amplitude or even the
sign of the supercurrent can be changed by the barrier strength,
the exchange field strength or the length of the $F_{y}$ layer, as
shown in Fig. \ref{f2}. The dependence of the supercurrent on the
barrier strength is because of the condition of resonance
transmission through double delta barriers
$sin(2kL_2)=-4U/(U^2+4)$ with $k$ the wave vector of particles.
Fig. \ref{h2l2} shows the anomalous supercurrent at zero phase
difference as functions of $h_2$ and $L_2$ for the $SF_xF_yF_zS$
junction. It is noted that the dependence on $h_2$ exhibits a
period of $2\pi/L_2=0.2$ which confirms the occurrence of
resonance transmission of electrons and holes through the $F_{y}$
layer. And the dependence on $L_2$ exhibits two periodic behavior.
One period is nearly $\pi$ and the other is $10\pi$. Because the
wavevector of "spin-up" electrons and holes is $k_{+}\approx k_{F}-\frac{h_2%
}{2}$ while the wavevector of "spin-down" electrons and holes is $%
k_{-}\approx k_{F}+\frac{h_2}{2}$ in the $F_y$ layer, the period $\pi$ stems
from $2\pi/2k_F=\pi$ with $k_F=1$. The other period $10\pi$ stems from $%
2\pi/2h_2=10\pi$ which is related to the difference of wavevectors $%
k_+-k_-=-h_2$.

\begin{figure}[btp]
\centering
\includegraphics[bb=18 14 305 222, width=3.405in]{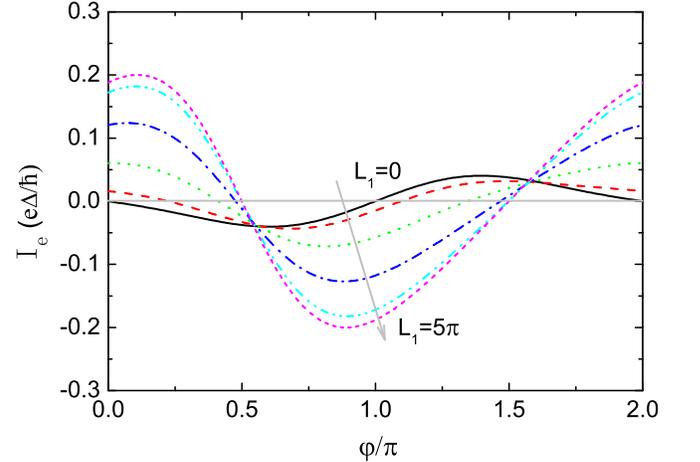}
\caption{(Color online) $I_e$ versus $\protect\varphi$ for the
$SF_xF_yF_zS$ junction with different $L_1$: varying $L_1$ from
$0$ to $5\protect\pi$ with a step $\protect\pi$, $L_3=L_1$. The
other parameters are chosen as: $U_0=2$, $h_1=h_2=h_3=0.1$,
$L_2=5\protect\pi$, $T/T_{c}=0.5$.} \label{l13}
\end{figure}

\begin{figure}[btp]
\centering
\includegraphics[bb=16 17 208 249, width=3.05in]{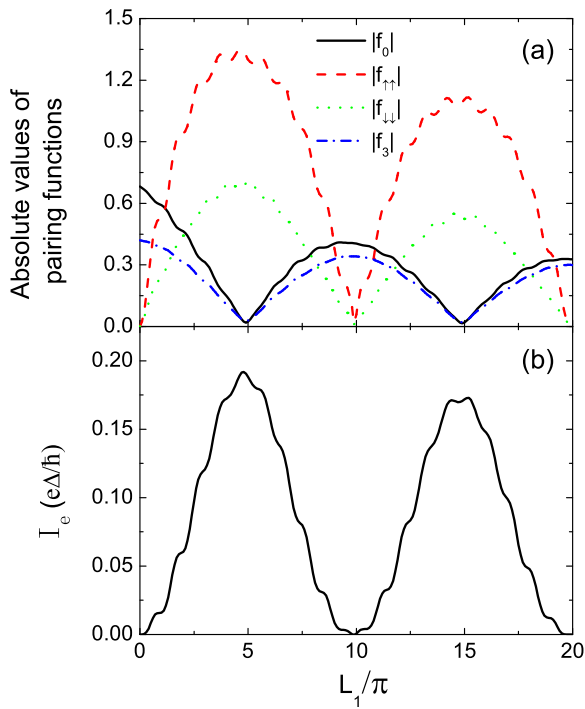}
\caption{(Color online) The absolute values of pair functions in the $F_y$
layer (a) and the anomalous supercurrent at zero phase difference (b) as
functions of the length of the $F_x$ and $F_z$ layers for the $SF_xF_yF_zS$
junction. $L_3=L_1$. The other parameters are the same as in Fig. \protect
\ref{l13}.}
\label{pair}
\end{figure}

If anyone of the two conditions $\phi _{1}=\phi _{3}=0$ and $%
h_{1}L_{1}=h_{3}L_{3}=(n+1/2)\pi $ is not satisfied, the pure equal-spin
triplet correlation in the $F_{y}$ layer is changed. For example, if we vary
the length or the exchange field strength of the $F_{x}$ and $F_{z}$ layers,
the spin precession angle of electron and hole in a round trip in the $F_{x}$
and $F_{z}$ layers is not $\pi $ any more. The reflected hole will have both
the same spin component and the opposite spin component to the incident
electron in the $F_{y}$ layer. Now the correlation is the mixing of singlet
and triplet. But only the triplet correlation can contribute to the
anomalous Josephson current, so the anomalous supercurrent is reduced with
increasing singlet component. Fig. \ref{l13} shows that both the amplitude
of the supercurrent and the equilibrium phase difference is tuned by the
length of the $F_x$ and $F_z$ layers. To study the characteristics of Cooper
pairs in the $F_{y}$ layer in detail, the pair function can be defined by
the anomalous Green function and be decomposed into four components \cite%
{eschrig07,tanaka07}%
\begin{equation}
\sum_{\omega _{n}>0}G_{\omega _{n}}^{eh}(x,x)=i\sum_{\nu =0}^{3}f_{\nu
}(x)\sigma _{\nu }\sigma _{2},  \label{pairing}
\end{equation}%
where $G_{\omega _{n}}^{eh}$ is the anomalous electron-hole correlation
function, $\sigma _{0}$ is the unit matrix and $\sigma _{\nu }(\nu =1,2,3)$
are three Pauli matrices. In Eq. ( \ref{pairing}) , the frequency summation
is only made over positive frequencies because the triplet pair functions
are odd functions of frequency. $f_{0}$ ($f_{3}$) is the pairing function of
spin-singlet (spin-triplet) pairs with spin structure of $\left[ \left\vert
\uparrow \downarrow \right\rangle -(+)\left\vert \downarrow \uparrow
\right\rangle \right] /\sqrt{2}$. The pairing functions of $\left\vert
\uparrow \uparrow \right\rangle $ and $\left\vert \downarrow \downarrow
\right\rangle $ pairs are given by $f_{\uparrow \uparrow }=if_{2}-f_{1}$ and
$f_{\downarrow \downarrow }=if_{2}+f_{1}$, respectively. Fig. \ref{pair} (a)
shows the absolute values of pairing functions at the center of the $F_{y}$
layer as functions of the length of the $F_{x}$ and $F_{z}$ layers for the $%
SF_{x}F_{y}F_{z}S$ junction. The equal-spin pair functions and opposite-spin
pair functions oscillate with the length of the $F_{x}$ and $F_{z}$ layers
which determines the angle of spin precession of quasiparticles in these two
layers. Compared with the anomalous supercurrent shown in Fig. \ref{pair}
(b), we can see that the anomalous supercurrent is nearly proportional to
the equal-spin triplet correlations.

\section{Conclusion}

In summary, we predict a tunable anomalous Josephson effect in $%
SF_{1}F_{2}F_{3}S$ junction where the three F layers have noncoplanar
magnetizations. The superconducting correlation can be completely triplet in
the $F_{2}$ layer due to the spin precession of electrons and holes in the $%
F_{1}$ and $F_{3}$ layers. If the condition $h_{1}L_{1}=h_{3}L_{3}=(n+1/2)%
\pi $ is satisfied, an electron incident to the left (right) superconductor
will precess its spin by $\frac{\pi }{2}$ in the $F_{1}$ ($F_{3}$) layer
before it arrives at the supercondcutor and the Andreev-reflected hole
proceeds to precess the spin by $\frac{\pi }{2}$ when it goes back to the $%
F_{2}$ layer. Thus the Andreev-reflected hole will have the same spin with
the incident electron and the complete triplet correlation arises in the $%
F_{2}$ layer. The two spin-resolved Andreev bound states carry two
spin-polarized supercurrents which have opposite phase shifts and
different amplitude thus leading to an anomalous Josephson
current. And the phase shift in the anomalous current-phase
relation is also a result of the spin precession of electron and
hole in the $F_{1}$ and $F_{3}$ layers. The equilibrium phase
difference of the anomalous supercurrent can be tuned by the
lengths, the exchange energies, and the magnetization orientations
of the $F_{1}$ and $F_{3}$ layers. And the amplitude of the
supercurrent can be tuned by the barriers between the F layers or
by the length and the exchange energy of the $F_{2}$ layer.

\begin{acknowledgments}
The work described in this paper was supported by the General Research Fund
of the Research Grants Council of Hong Kong SAR, China under Project No.
CityU 100308/08P.
\end{acknowledgments}

\end{document}